\catcode`\@=11  % This allows us to modify PLAIN macros.
\def\jnl{f} % specify  t,f,i or n
\def\revthree{t }    % t = revtex3
\def\revfour{f }     % f = revtex4
\def\iop{i }         % i = iop style
\def\nojnl{n }       % n = no style files (latex2)

\if\jnl\iop
 \documentclass[12pt]{iopart}
 \usepackage{epsf}  % Include figure files
\fi

\if\jnl\revfour
 \documentclass[aps,draft,twocolumn,eqsecnum]{revtex4}
 \usepackage{epsf}  % Include figure files
\fi

\if\jnl\revthree
\documentstyle[aps,eqsecnum,prd,epsf]{revtex}
\fi

\if\jnl\nojnl
\documentstyle[12pt]{article}
\fi
\newtheorem{theorem}{Theorem}

\newtheorem{prop}{Proposition}%[section]

\begin{document}
% \baselineskip 17pt

%-----------------------------------------------------------------------
\title{
Diagonalizability of Constraint Propagation Matrices}
%-----------------------------------------------------------------------
% revtex4 style --------------------------------------------------------
\if\jnl\revfour
\author{Gen Yoneda}\email{yoneda@aoni.waseda.jp}
\author{Hisa-aki Shinkai}\email{hshinkai@postman.riken.go.jp}
\affiliation{
${}^\ast$ Department of Mathematical Sciences, Waseda University,
Okubo, Shinjuku, Tokyo,  169-8555, Japan
\\
${}^\dagger$ Computational Science Division
Institute of Physical \& Chemical Research (RIKEN), \\
Hirosawa, Wako, Saitama, 351-0198 Japan
}
\fi

% revtex3 style --------------------------------------------------------
\if\jnl\revthree
\author{Gen Yoneda ${}^\dagger$ and Hisa-aki Shinkai ${}^\ddagger$}
\address{
{\tt yoneda@aoni.waseda.jp} ~~~ {\tt hshinkai@postman.riken.go.jp}
\\
${}^\dagger$ Department of Mathematical Sciences, Waseda University,
Okubo, Shinjuku, Tokyo,  169-8555, Japan
\\
${}^\dagger$ Computational Science Division,
Institute of Physical \& Chemical Research (RIKEN), \\
Hirosawa, Wako, Saitama, 351-0198 Japan
}
\fi

%-----------------------------------------------------------------------
%\date{September 28, 2002 ~~~ gr-qc/0209106}
%\date{Jordan version 5 ~~ HS 2002/10/22}
%\date{Revised version ~~ 2002/10/22 ~~ gr-qc/0209106 v2}
\date{2002/12/23 ~~ gr-qc/0209106 ~~ to appear in Class. Quant. Grav. Lett.}
%\date{\today  ~~ draft version 10 ~~ HS }
%-----------------------------------------------------------------------
\begin{abstract}
%=======================================================================
%<<<<<<<<<<<<< ABSTRACT >>>>>>>>>>>>>>>%
%=======================================================================
In order to obtain stable and accurate general relativistic simulations,
re-formulations of the Einstein equations are necessary.
In a series of our works, we have proposed using eigenvalue analysis of
constraint propagation equations for evaluating violation behavior of
constraints.
In this article, we classify asymptotical behaviors of constraint-violation
into three types
(asymptotically constrained, asymptotically bounded, and diverge),
and give their necessary and sufficient conditions.
We find that degeneracy of eigenvalues sometimes leads constraint evolution
to diverge (even if its real-part is not positive), and conclude that
it is quite useful to check the diagonalizability of
constraint propagation matrices.
The discussion is general and can be applied to any numerical treatments of
constrained dynamics.
\end{abstract}
 \pacs{04.20.-q, 04.20.Fy, 04.25.-g and 04.25.Dm}
%\multicols{2}

%\keywords{General Relativity, Numerical Relativity}
%-----------------------------------------------------------------------
\maketitle

%=======================================================================
%%%%%%%%%%%%%%%%%%%%%%%%%%%%%%%%%%%%%%%%%%%%%%%%%%%%%%%%%%%%%%%%%%%%%%
\section{Introduction}
%%%%%%%%%%%%%%%%%%%%%%%%%%%%%%%%%%%%%%%%%%%%%%%%%%%%%%%%%%%%%%%%%%%%%%

So-called numerical
relativity  (computational simulations in general relativity) is
a promising research field
%matching with
having implications for
ongoing
astrophysical observations such as gravitational wave astronomy
\cite{NRreviews}.
Many simulations of binary compact objects
have revealed that mathematically
equivalent sets of evolution equations show different numerical
stability in the free-evolution scheme.

There are many approaches to re-formulate the Einstein equations
for obtaining a longterm stable and accurate numerical evolution
(e.g. see references in  \cite{novabook}).
In a series of our works, we have proposed
%to construct
the construction of
a system that has its constraint surface as an attractor.
By applying eigenvalue analysis of constraint propagation equations,
we showed that there {\it is} a constraint-violating mode
in the standard Arnowitt-Deser-Misner (ADM)
evolution system \cite{ADM,ADM-York} when it is applied to
a single non-rotating black-hole space-time\cite{adjADM2}.
We also found that such a constraint-violating mode can be
%killed
compensated for
if we adjust the evolution equations with a particular
modification using constraint terms like the one proposed
by Detweiler \cite{detweiler}.

Our predictions are borne out in simple
numerical experiments using the Maxwell, Ashtekar, and ADM systems
\cite{ronbun2,adjADM,adjADM2,adjBSSN}.
There are also several numerical experiments to confirm our predictions
are effective\cite{pennstate,illinois}.
However we have not yet obtained definite guidelines for specifying
the above adjusting terms and their multipliers.

In this article, we show the essential steps in analyzing
constraint amplification factors (defined in \S \ref{secCAFdef}).
In \S \ref{sec3},
we show that degeneracy of eigenvalues sometimes leads
constraint evolution to diverge.  This observation suggests the
importance of checking the diagonalizability of characteristic matrices,
and gives further insights for constructing an asymptotically
constrained system.

% The discussion is general and the idea can be applied to any
% numerical treatments of constrained dynamics.

%%%%%%%%%%%%%%%%%%%%%%%%%%%%%%%%%%%%%%%%%%%%%%%%%%%%%%%%%%%%%%%%%%%%%%
\section{A guideline to obtain a robust evolution system}
%%%%%%%%%%%%%%%%%%%%%%%%%%%%%%%%%%%%%%%%%%%%%%%%%%%%%%%%%%%%%%%%%%%%%%
\subsection{Idea of Adjusted system}
We begin by reviewing our proposal for an ``adjusted system".

Suppose we have a dynamical system of variables $u^a (x^i,t)$,
which has evolution equations,
\begin{equation}
\partial_t u^a = f(u^a, \partial_i u^a, \cdots), \label{ueq}
\end{equation}
and the (first class) constraints,
\begin{equation}
C^\alpha (u^a, \partial_i u^a, \cdots) \approx 0.
\end{equation}
Note that we do not require (\ref{ueq}) to form a first-order hyperbolic
form.
We propose  to investigate
the evolution equation of $C^\alpha$ (constraint propagation),
\begin{equation}
\partial_t C^\alpha = g(C^\alpha, \partial_i C^\alpha, \cdots),
 \label{Ceq}
\end{equation}
for evaluating violation features of constraints.

%----------------------------------------------------------
The character of constraint propagation, (\ref{Ceq}),
will vary when we modify the original evolution equations.
Suppose we modify (adjust) (\ref{ueq}) using constraints
\begin{equation}
\partial_t u^a = f(u^a, \partial_i u^a, \cdots)
+ F(C^\alpha, \partial_i C^\alpha, \cdots), \label{DeqADJ}
\end{equation}
then (\ref{Ceq}) will also be modified as
\begin{equation}
\partial_t C^\alpha = g(C^\alpha, \partial_i C^\alpha, \cdots)
+ G(C^\alpha, \partial_i C^\alpha, \cdots). \label{CeqADJ}
\end{equation}
Therefore, finding a proper adjustment
$F(C^\alpha, \cdots)$ is a quite important problem.

Hyperbolicity analysis  may be a way to
evaluate constraint propagation, (\ref{Ceq}) and (\ref{CeqADJ}) \cite{Fri-con}.
However, this requires  (\ref{Ceq}) to be a first-order system
which is easy to be broken.
(See e.g. Detweiler-type adjustment \cite{detweiler} in the
ADM formulation \cite{adjADM2}).
Furthermore hyperbolicity analysis only concerns the principal part of the
equation, that may fail to analyze the detail evaluation of evolution.

Alternatively, we have proceeded
an eigenvalue analysis of the whole RHS in
(\ref{Ceq}) and (\ref{CeqADJ}) after a suitable homogenization,
which may compensate for the above difficulties of hyperbolicity analysis.

%----------------------------------------------------------
%----------------------------------------------------------
\subsection{CP matrix and CAF} \label{secCAFdef}
%----------------------------------------------------------
%----------------------------------------------------------

%--------------------------------------------------------  BOX begins ---
We propose to transform the constraint propagation
equation,  (\ref{Ceq}) and (\ref{CeqADJ}),  into Fourier modes,
\begin{eqnarray}
&& \partial_t \hat{C}^\alpha = \hat{g}(\hat{C}^\alpha)
=M^\alpha{}_{\beta} \hat{C}^\beta,
\nonumber \\ &&
\mbox{~where~}
%\lambda(x,t)^\alpha
%=\displaystyle{\int} \hat{\lambda}(k,t)^\alpha\exp(ik\cdot x)d^3k, \quad
C(x,t)^\alpha
=\displaystyle{\int} \hat{C}(k,t)^\alpha\exp(ik\cdot x)d^3k,
\label{CeqF}
\end{eqnarray}
then to analyze the eigenvalues, say $\Lambda_\alpha$,  of the
coefficient matrix, $M^\alpha{}_{\beta}$.
We call $\Lambda_\alpha$ and $M^\alpha{}_{\beta}$ the constraint
amplification factors (CAFs)
and constraint propagation matrix (CP matrix), respectively.
%--------------------------------------------------------  BOX ends ---

So far we have proposed the following heuristic conjectures
\cite{ronbun2,adjADM,adjADM2,adjBSSN}:
%--------------------------------------------------------  BOX begins ---
\begin{itemize}
\item[(A)] If the CAF has
a {\it negative real-part }
(the constraints  are necessarily
%forced to be
diminished), then we see
more stable evolution than a system which has a positive CAF.
\item[(B)] If the CAF has a {\it non-zero  imaginary-part }
(the constraints are propagating away), then we see
more stable evolution than a system which has
a zero CAF.
\end{itemize}
%--------------------------------------------------------  BOX ends ---
%We found that the system becomes more stable when
%more $\Lambda$s satisfy the above criteria.
We
observe
%% remark
that this eigenvalue analysis requires the
fixing of a particular background space-time,
since the CAFs depend on the dynamical variables, $u^a$.

%----------------------------------------------------------
%----------------------------------------------------------
\subsection{Classification of Constraint propagations}
%----------------------------------------------------------
%----------------------------------------------------------
The CAFs indicate the evolution of constraint violations
(definitely its Fourier modes).
It is natural to assume that a divergence of constraint
norm is related to the numerical blow-ups.
Therefore we classify the fundamental evolution property of
constraint propagation equation (\ref{CeqF}) as follows:
%--------------------------------------------------------  BOX begins ---
\begin{itemize}
\item[(C1)]  {\it Asymptotically constrained : }
Violation of constraints decays (converges to zero).
\item[(C2)]  {\it Asymptotically bounded : }
Violation of constraints is bounded at a certain value.
\item[(C3)]  {\it Diverge : } At least one constraint will diverge.
\end{itemize}
%--------------------------------------------------------  BOX ends ---
Note that (C1) $\subset$ (C2).
We will derive the necessary and sufficient conditions for (C1) and
(C2)  in the next section.

%%%%%%%%%%%%%%%%%%%%%%%%%%%%%%%%%%%%%%%%%%%%%%%%%%%%%%%%%%%%%%%%%%%%%%
\section{Conditions for (C1) and (C2)} \label{sec3}
%%%%%%%%%%%%%%%%%%%%%%%%%%%%%%%%%%%%%%%%%%%%%%%%%%%%%%%%%%%%%%%%%%%%%%
%=======================================================================
\subsection{Preparation}
Hereafter, we consider a set of evolution equations,
\begin{eqnarray}
\partial_t  C^i (t)= M^i{}_j C^j,
\label{CP}
\end{eqnarray}
where $C^i ~ (i=1, \cdots, n)$ is a complex-valued vector,
$M^i{}_j$ is a $n \times n$
complex-valued matrix, and
$C^i(t)$ is assumed to have finite-valued initial data $C^i(0)$.

Without loss of generality,
the CP matrix $M$ can be assumed to be a Jordan normal form, since
within complex-valued operations
all the matrices can be converted to this form.
Suppose that $M$ has $r$ different eigenvalues
$(\lambda_1,\cdots,\lambda_r)$, where  $r\leq n$.
Let the multiplicity of $\lambda_k$
as $n_k$, where $\sum_{k=1}^r n_k=n$.
$M$ can be expressed as
\begin{eqnarray}
M&=&J_1\dot{+}\cdots \dot{+}J_r
:=\left(\matrix{
J_1 &   &O \cr
& \ddots  & \cr
O & & J_r }
\right),
\end{eqnarray}
where the cell size of $J_k$ is $n_k \times n_k$.
The Jordan matrix $J_k$ is then expressed using
a Jordan block $J_{k\ell}$,
\begin{eqnarray}
J_k&=&J_{k1}\dot{+}\cdots \dot{+}J_{km},
\label{jordanblock}
\\
J_{k\ell}&:=&
\left(\matrix{
\lambda_k & 1 &  &O \cr
& \ddots & \ddots & \cr
O & & \lambda_k & 1 }
\right).
\end{eqnarray}
Note that  $J_{k\ell}$ is $n_{k\ell} \times n_{k\ell}$, 
$\sum_{\ell=1}^m n_{k\ell}=n_k$,
$m=n- \mbox{rank} (M-\lambda_k E)$
and $\max_{\ell} (n_{k\ell})=\nu_k$. 
The minimum polynomial of $M$ is written as
\begin{equation}
\mu_M(t)=(t-\lambda_1)^{\nu_1} \cdots (t-\lambda_r)^{\nu_r}.
\end{equation}
If $J_k$ is diagonal
(i.e. $n_k = n- \mbox{rank} (M-\lambda_k E)$), then $\nu_k=1$
for that $k$.
If $M$ is diagonalizable
(i.e. $n_k = n- \mbox{rank} (M-\lambda_k E)$ for $\forall k$), then
$\nu_k=1$ for all $k$.

We then have the following statement.
%=========================== %>>>>>>>>>>>>>>>>>>>>>>>>>> PROP A
\begin{prop}
%Suppose $M$ is a triangular matrix,
The solution of %(\ref{CP2})
\begin{equation}
\partial_t C_a=J_k C_a
\label{CP2}
\end{equation}
can be expressed formally as
\begin{eqnarray}
C_a(t)=
\exp(\lambda_k t)
\sum_{\ell=0}^{\nu_k-1} (a^{(k)}_\ell t^\ell).
\label{kino}
\end{eqnarray}
%where $\lambda_j$ is the corresponding eigenvalues of (\ref{CP2})
%(i.e. $M_{ii}=\lambda_j$),
%and
%$\mu_i$ is the multiplicity of $\lambda_i$ up to $i \leq j$.
%$\mu_j$ is the power of $(t-\lambda_j)$
%in the minimal polynomial $\mu_{M}(t)$.
\label{propA}
\end{prop}
%=========================== %>>>>>>>>>>>>>>>>>>>>>>>>>> PROP A
%We show a proof (by mathematical induction) in Appendix \ref{AppA}.
%{}From (\ref{kino}),
A proof is available by mathematical induction.
Suppose that $J_{k1}$ is $\nu_k\times\nu_k$ 
which is the maximal size $J_k$.
By direct calculation,
we have that 
$\partial_t C_a=J_{k1} C_a$ yields
(\ref{kino}) with $t$-polynomial of degree $(\nu_k-1)$.
Then we see that (\ref{kino}) is satisfied
in general.

{}From this proposition,
the highest power $N_k$ in $t-$polynomial in (\ref{kino})
is bounded by
$0 \leq N_k \leq
\nu_k -1$.
The matrix $J_k$ in (\ref{CP2}) can be directly extended to the full
CP matrix, $M$,  in (\ref{CP}).
Therefore the highest power $N$ in all constraints
is bounded by
\begin{equation} 0 \leq N
%\leq \max_{1 \leq j \leq r}\mu_j-1
\leq
%\max_{1 \leq j \leq r}(\mbox{multiplicity of }\lambda_j)-1.
\max_{1 \leq k \leq r}(n_k)-1.
\end{equation}
%%%%%%%%%%%%%%%%%%%%%%%%%%%%%%%%%%%%%%%%%%%%%%%%%%%%%%%%%%%%%%%%%%%%%%
\subsection{Asymptotically Constrained CP}
The following Propositions \ref{prop1} and \ref{prop2} give us
the next theorem.
%=========================== %>>>>>>>>>>>>>>>>>>>>>>>>>> Theorem 1
\begin{theorem}
{Asymptotically constrained evolution}
(violation of constraints converges to zero) is
obtained if and only if all the real parts of the CAFs are negative.
\label{theoremC1}
\end{theorem}
%=========================== %>>>>>>>>>>>>>>>>>>>>>>>>>> Theorem 1

%=========================== %>>>>>>>>>>>>>>>>>>>>>>>>>> PROP 1
\begin{prop}
All the real part of CAFs are negative
$\Rightarrow$ Asymptotically constrained evolution.
\label{prop1}
\end{prop}
%=========================== %>>>>>>>>>>>>>>>>>>>>>>>>>> PROP 1
proof) We use the expression
(\ref{kino}).  If
$ \Re e (\lambda_k)<0$ for $\forall k$, then $C_i$ will converge to zero at
$t\to \infty$ no matter what the $t-$polynomial terms are.
${\kern1pt\vbox{\hrule height 1.2pt\hbox{\vrule width1.2pt
\hskip 3pt\vbox{\vskip 6pt}\hskip 3pt\vrule width 0.6pt}
\hrule height 0.6pt}
\kern1pt}$

%=========================== %>>>>>>>>>>>>>>>>>>>>>>>>>> PROP 2
\begin{prop}
Asymptotically constrained evolution $\Rightarrow$
All the real parts of the CAFs are negative.
\label{prop2}
\end{prop}
%=========================== %>>>>>>>>>>>>>>>>>>>>>>>>>> PROP 2
proof)
We show the contrapositive.
Suppose there exists an eigenvalue $\lambda_1$ of
%such as
%$\re(\lambda_1)\geq 0 $.
which the real-part is non-negative.
%By setting $\lambda_1$  at
%the lower-end of the triangular matrix $M$ in (\ref{CP2}),
Then we get
$\partial_t C_1=\lambda_1 C_1$  of which the solution is
$C_1=C_1(0)\exp(\lambda_1 t)$. $C_1$ does not converge to zero.
${\kern1pt\vbox{\hrule height 1.2pt\hbox{\vrule width1.2pt
\hskip 3pt\vbox{\vskip 6pt}\hskip 3pt\vrule width 0.6pt}
\hrule height 0.6pt}
\kern1pt}$

%%%%%%%%%%%%%%%%%%%%%%%%%%%%%%%%%%%%%%%%%%%%%%%%%%%%%%%%%%%%%%%%%%%%%%
\subsection{Asymptotically Bounded CP}
The following Propositions \ref{prop3} and \ref{prop4} give us
the next theorem.
%=========================== %>>>>>>>>>>>>>>>>>>>>>>>>>> THEOREM 2
\begin{theorem}
{ Asymptotically bounded evolution}
(all the constraints are bounded at a certain value) is
obtained if and only if
all the real parts of CAFs are not positive
and $J_k$ is diagonal
%$n_k = n-\mbox{rank}(M-\lambda_k E)$
when $\Re e (\lambda_k)=0$.
\label{theoremC2}
\end{theorem}
%=========================== %>>>>>>>>>>>>>>>>>>>>>>>>>> THEOREM 2

%=========================== %>>>>>>>>>>>>>>>>>>>>>>>>>> Corollary
\noindent{\bf Corollary}
{\it
Asymptotically bounded evolution
is obtained if
the real parts of CAFs are not positive
and the CP matrix $M^\alpha{}_\beta$ is diagonalizable. }
%=========================== %>>>>>>>>>>>>>>>>>>>>>>>>>> Corollary

%=========================== %>>>>>>>>>>>>>>>>>>>>>>>>>> PROP 3
\begin{prop}
All the real parts of CAFs are not positive
and $J_k$ is diagonal
%$n_k = n-\mbox{rank}(M-\lambda_k E)$
when $\Re e (\lambda_k)=0$
$\Rightarrow$
Asymptotically bounded evolution.
\label{prop3}
\end{prop}
%=========================== %>>>>>>>>>>>>>>>>>>>>>>>>>> PROP 3

proof)
We use the expression (\ref{kino}).
When $\Re e (\lambda_k)<0$,
$\exp(\lambda_k t) \times (t \mbox{-polynomials})$ will converge to zero
no matter what the $t-$polynomial terms are.
When $\Re e (\lambda_k)=0$,
we see $\nu_k=1$ from the assumption of diagonality of $J_k$.
%$n_k = n-\mbox{rank}(M-\lambda_k E)$.
So we see the $t-$polynomial terms are constant
and $\exp(\lambda_k t)$ is bounded.
${\kern1pt\vbox{\hrule height 1.2pt\hbox{\vrule width1.2pt
\hskip 3pt\vbox{\vskip 6pt}\hskip 3pt\vrule width 0.6pt}
\hrule height 0.6pt}
\kern1pt}$

%=========================== %>>>>>>>>>>>>>>>>>>>>>>>>>> PROP 4
\begin{prop}
Asymptotically bounded evolution
$\Rightarrow$
All the real parts of the CAFs are not positive
and $J_k$ is diagonal
%$n_k= n-\mbox{rank}(M-\lambda_k E)$
when $\Re(\lambda_k)=0$.
\label{prop4}
\end{prop}
%=========================== %>>>>>>>>>>>>>>>>>>>>>>>>>> PROP 4

proof) We show the contrapositive.
If there exists an eigenvalue of which the real-part is positive, then
constraints will diverge
no matter what the $t-$polynomial terms are.
Therefore we try to show that constraints will diverge
when all the real-parts of eigenvalues are non-positive,
and there exists $\lambda_k$ such that
$\Re(\lambda_k)=0$ and its Jordan matrix $J_k$ is not diagonal.
%$n_k \neq n-\mbox{rank}(M-\lambda_k E)$ for $k$ that gives
%$\Re(\lambda_k)=0$.

{}
%From the assumption
%of $n_k \neq n-\mbox{rank}(M-\lambda_k E)$,
Since Jordan matrix $J_k$ is not diagonal,
 we see the power of $t-$polynomial $\nu_k$ is greater than 1
in the expression (\ref{kino}).
%(See Appendix in detail.)
Thus we have that
(\ref{kino}) will diverge in $t \rightarrow \infty$.
%even if $\Re(\lambda_k)=0$.
${\kern1pt\vbox{\hrule height 1.2pt\hbox{\vrule width1.2pt
\hskip 3pt\vbox{\vskip 6pt}\hskip 3pt\vrule width 0.6pt}
\hrule height 0.6pt}
\kern1pt}$

%=======================================================================
\section{Concluding Remarks}
%=======================================================================
Two theorems
%can be summarized as Table \ref{TableClassification},
%and this
will give us a guideline to analyze a constraint-violating
mode of the system.
The result supports our previous heuristic conjecture (A), but also
suggests an ill-behaving case when CAFs are degenerated and its
real-part is zero, when the associated Jordan matrix is not diagonal.
This indicates the
importance of checking  the diagonalizability of constraint
propagation matrix $M$.

\begin{figure}[b]
\unitlength 1mm
\begin{picture}(85,75)
\put(-5,00){\epsfxsize=90mm \epsffile{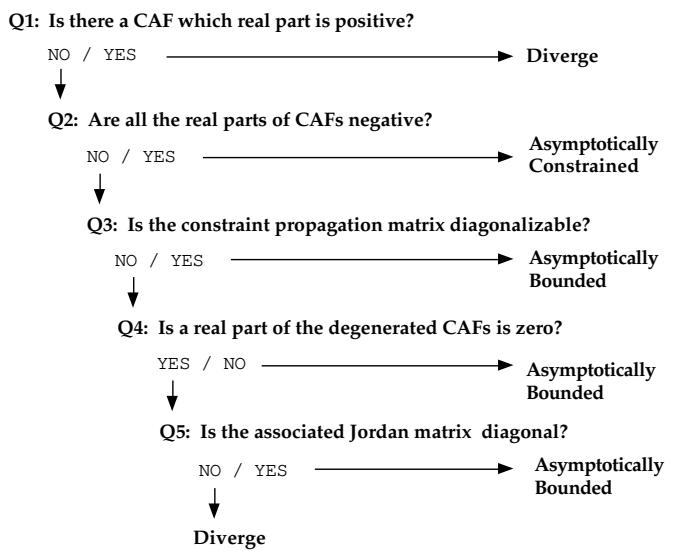} }
\end{picture}
\caption{A flowchart to classify the fate of constraint propagation. }
\label{fig:flowchart}
\end{figure}

Along the line of our evaluation of constraint propagation
equations (\ref{CP}), we propose a practical procedure
for this classification in Figure \ref{fig:flowchart}.
We think that this diagram will provide systematic predictions
for obtaining a robust evolution system in any constrained dynamics.

% \input{diagCP_fig}
%\begin{widetext}
%\begin{center}
%----------------------------------------------  Figure .{fig:flowchart}
%----------------------------------------------  Figure .{fig:flowchart}
%\end{center}
%\end{widetext}

The present classification is only on the fixed background
spacetime and only for $t\rightarrow \infty$.
It is still not clear at what value the constraints are bounded
if a limiting value exists.
Thus further modifications are underway.
We are also applying the present classification scheme to
various adjusted systems
of the Einstein equations (adjusted ADM, and further modified versions),
together with numerical experiments.  We hope to report on them 
in the near future.

The current constraint analysis only concentrates to the evolution equations
and does not include the effect of the boundary treatments. 
Since the eigenvalues are evaluated locally, it will be possible to include
the effect of numerical boundary conditions if they are expressed apparently
in a part of the evolution equations.  This is also the one direction to proceed
our future research.  Meanwhile, we would like to remark that one of our
proposed adjustments in \cite{adjBSSN} contributes to enforce the 
computational ability of the black-hole excision boundary treatment
\cite{illinois}.

By extending the notion of
 ``norm" or ``compactness" of constraint violations,
it might be interesting to define a new measure which monitors
a ``distance" between the constraint surface and an evolution sector in
constraint dynamics.

%====================================================================
\section*{Acknowledgements}
%====================================================================
HS thanks the Caltech Visitors Program for the Numerical Simulation of
Gravitational Wave Sources for their hospitality, 
%when preparing the revised version of this manuscript.
where a portion of this work was completed.
HS is supported by the special postdoctoral researchers' program
at  RIKEN.
This work was supported partially by the Grant-in-Aid for Scientific
Research Fund of Japan Society of the Promotion of Science, No. 14740179.

%%%%%%%%%%%%%%%%%%%%%%%%%%%%%%%%%%%%%%%%%%%%%%%%%%%%%%%%%%%%%%%%%%%%%%
%234567890123456789012345678901234567890123456789012345678901234567890
%000000001111111111222222222233333333334444444444555555555566666666667
%%%%%%%%%%%%%%%%%%%%%%%%%%%%%%%%%%%%%%%%%%%%%%%%%%%%%%%%%%%%%%%%%%%%%%

%====================================================================
%-------------------------------------------------- references ------
%====================================================================

% \input{diagCP_fig}

%%%%%%%%%%%%%%%%%%%%%%%%%%%%%%%%%%%%%%%%%%%%%%%%%%%%%%%%%%%%%%%%%%%%%%
% end of TeX file
%   Diagonalizability of Constraint Propagation Matrix
%
%%%%%%%%%%%%%%%%%%%%%%%%%%%%%%%%%%%%%%%%%%%%%%%%%%%%%%%%%%%%%%%%%%%%%%
\end{document}